\documentclass[aps,pre,reprint, amsmath, amssymb,superscriptaddress]{revtex4-1}

\usepackage{tikz}
\usetikzlibrary{arrows.meta}
\usepackage{bm}
\newcommand{\beq}{\begin{equation}}
\newcommand{\eeq}{\end{equation}}

\usepackage[retainorgcmds]{IEEEtrantools}
\usepackage{graphicx,tikz,placeins}
\usepackage{mathrsfs}
\usepackage{amsmath,amssymb,amsfonts,physics}
\usepackage{color}
\usepackage{float}
\usepackage{times,txfonts}
\usepackage{nicefrac}
\usepackage[colorlinks=true,linkcolor=blue,urlcolor=blue,citecolor=blue,pdfusetitle]{hyperref}
\usepackage{physics}
\usepackage{soul}

\usepackage[cp1252,utf8]{inputenc}

\begin{document}

\title{Thermodynamics of a collisional quantum-dot machine: the role of stages.}

\author{C. E. Fernández Noa}
\affiliation{Universidade de São Paulo,
Instituto de Física,
Rua do Matão, 1371, 05508-090
São Paulo, SP, Brasil}
´
\affiliation{UHasselt, Faculty of Sciences, Theory Lab, Agoralaan, 3590 Diepenbeek, Belgium}
\author{C. E. Fiore}
\author{F. F.  S. Filho}
\affiliation{Universidade de São Paulo,
Instituto de Física,
Rua do Matão, 1371, 05508-090
São Paulo, SP, Brasil}
\author{B. Wijns}
\author{B. Cleuren}
\affiliation{UHasselt, Faculty of Sciences, Theory Lab, Agoralaan, 3590 Diepenbeek, Belgium}
\date{\today}
\begin{abstract}

Sequential (or collisional) engines have been put forward as an alternative candidate for the realisation of reliable engine setups.  Despite this, the role of the different stages and the influence of the intermediate reservoirs is not well understood. We introduce the idea of conveniently adjusting/choosing  intermediate reservoirs at engine devices as a strategy for optimizing its performance. This is done by considering a minimal model composed of a quantum-dot machine sequentially exposed to distinct reservoirs at each stage, and for which thermodynamic quantities (including power and efficiency) can be obtained exactly from the framework of stochastic thermodynamics, irrespective the number of stages. Results show that a significant gain can be obtained by increasing the number of stages and conveniently choosing their parameters.
\end{abstract}
\maketitle
\section{Introduction}
Stochastic engines are devices that convert a given amount of energy, say heat\textbf{,} into work or vice-versa. In contrast to  macroscopic engines, they operate  at the nanoscale and consequently the relevant thermodynamic quantities are subjected to fluctuations at the microscopic level, above all in power and efficiency. Although an ideal engine is always desired to operate at high power, high efficiency and low (power) fluctuations, these conditions can never be satisfied simultaneously. For this reason the development of distinct approaches/trade-offs has been strongly
levered in the last years, such as by including the variation
of external parameters \cite{PhysRevE.61.4774}, cyclic operations under quasistatic conditions \cite{PhysRevLett.121.120601}, interaction between particles \cite{gatien,mamede2021obtaining}, dynamics based on control via shortcuts
 to adiabaticy \cite{RevModPhys.91.045001,deffner2020thermodynamic,pancotti2020speed}, to isothermality \cite {2206.02337}, maximization of power \cite{schmiedl2007efficiency, cleuren2015universality, van2005thermodynamic, esposito2010quantum, seifert2011efficiency, izumida2012efficiency, golubeva2012efficiency, holubec2014exactly, bauer2016optimal,tu2008efficiency, ciliberto2017experiments,bonanca2019,rutten2009reaching} and efficiency \cite{karel2016prl,mamede2021obtaining} and more recently the strategies based on
synchronized operation  under ordered arrangements
 \cite{forao2023powerful} or Pareto optimal cycles for power, efficiency and fluctuations \cite{erdman2022pareto}.

Sequential (or collisional) engines have been put forward as an alternative candidate for the realisation of reliable  thermal  engines   \cite{rosas1,rosas2} and novel engine setups \cite{stable2020thermodynamics,noa2021efficient,fernando,pedro}. They consist of sequentially placing the system/engine in contact with distinct thermal reservoirs and subjected to external driving forces during each stage (stroke) of the cycle. Each stage is characterized by the connected thermal reservoir. The time needed to switch between the thermal baths at the beginning/end of each stage is neglected. Despite its reliability in distinct situations \cite{benn1,maru,saga,parrondo}, the conditions to be imposed in order to provide  a better performance have not been fully understood and for this reason distinct (and recent) approaches for enhancing its performance have been proposed and analyzed. Among them, we cite the convenient choices of the duration of each stroke \cite{pedro,noa2021efficient} and driving \cite{fernando,mamede2022obtaining}.

In this contribution, we address a less explored strategy for improving the performance of collisional engines: the number of stages and the reservoir parameters for each stage. The system we consider is a particle pump model introduced in references \cite{rosas1,rosas2}, consisting of a two-level system  
 sequentially brought into contact with distinct reservoirs allowing for the exchange of particles among reservoirs and the generation of a power output.  Quantum dot devices are one of
 the most prominent  system in the realm of stochastic and quantum thermodynamics, as in theoretical
 \cite{vsubrt2007exact,chvosta2010thermodynamics,verley2013modulated,harunari2020} and experimental \cite{khan2021efficient} studies. Due to  its simplicity, it presents several advantages such as an exact solution irrespective of the number of strokes and model parameters \cite{vsubrt2007exact,chvosta2010thermodynamics,verley2013modulated,harunari2020}.
 And so it provides full access to all relevant quantities. Another advantage  concerns that they can be projected to function either as heat or pump engines rather than Brownian engines,  which only can be operated as work-to-work
converters depending on the kind of external driving used as the work source \cite{stable2020thermodynamics,noa2021efficient,fernando,mamede2021obtaining}.
 A careful analysis over the space of parameters for distinct intermediate
 stages reveals that a remarkable gain can be obtained by increasing the number of stages and a suitable choice of parameters.

This paper is structured as follows: In Sec. II, the thermodynamic considerations
are derived irrespective of the number of strokes and exemplified for distinct 
stages. The engine performance when the stages are varied is investigated
in Sec. III and conclusions are drawn in Sec. IV.


\begin{figure}\label{fig1}
    \centering
\includegraphics[width=0.6\linewidth]{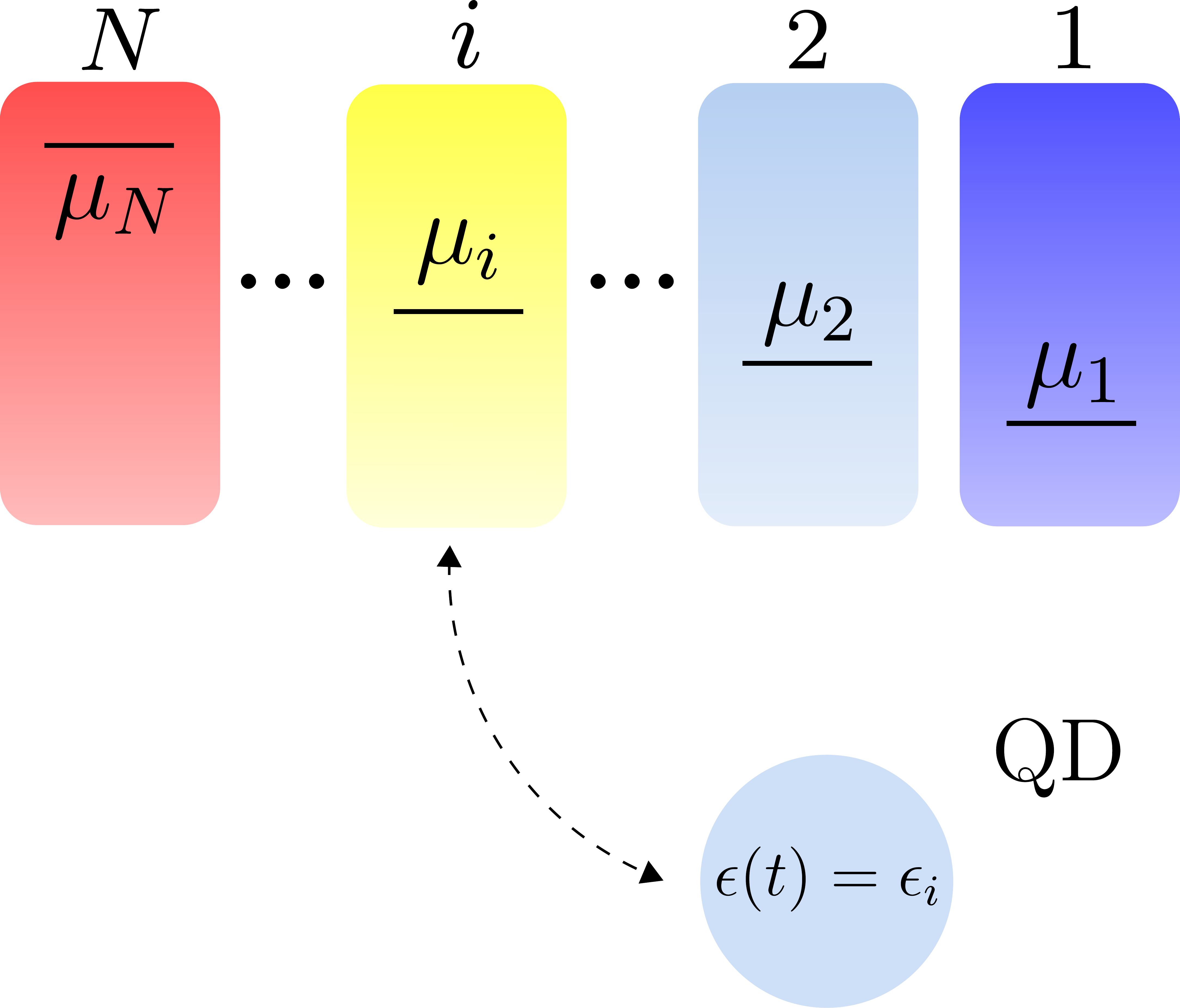}
    \caption{Sketch of a Quantum-dot setup  sequentially exposed to $N$ distinct thermal baths, each one at the interval $\tau_{i-1}\le t<\tau_i=i\tau/N$ characterized by chemical
    potential $\mu_i$ and temperature $T_i=T$.}
\end{figure}

\section{Collisional engines}
The model we consider consists of a quantum dot (QD) and $N$ thermal reservoirs. During operation, the QD is connected sequentially to each of the
reservoirs, and this for a time duration $\tau/N$. The total time to complete one cycle is $\tau$. The QD has a single energy level which is either filled or empty. At stage $i$ when the QD is connected to reservoir $i$ we write $p_i(t)$ for the probability for the QD to be filled. The cycle is started at time $t=0$ and the QD is connected to reservoir 1 which is characterised by the temperature $T_1$ and chemical potential $\mu_1$. The energy level of the QD is set to $\epsilon_1$. After a time duration $\tau/N$, it is disconnected and reconnected to reservoir 2 with $\mu_2$ and $T_2$. Furthermore, the energy level is changed from $\epsilon_1$ to $\epsilon_2$. This process of connecting/deconnecting is repeated until the $N$-th stage, marking a full cycle of period $\tau$, after which the process starts all over. For simplifying matters we set all $T_i=T$ so that we obtain an isothermal operation.
Figure~\ref{fig1} sketches  the actual setup.  For each stage $i$, we distinguish two \emph{phases}:
\begin{itemize}
    \item an \emph{exchange} phase during which the QD is connected to reservoir $i$ (chemical potential $\mu_i$). During this phase the energy level of the QD is $\epsilon_i$.
    \item an \emph{external driving} phase during which the QD is uncoupled from any reservoir and its energy level is changed from $\epsilon_i$ to $\epsilon_{i+1}$. Since the duration of this driving phase is irrelevant for the thermodynamic discussion we effectively set it equal to zero, so that the energy level change is instantaneously.
\end{itemize}
The time evolution of $p_i(t)$ at the $i$-th stroke is determined by the master equation
\begin{equation}\label{eq:master}
\dot{p_i}(t)=\left[1-p_i(t)\right]\omega^{(i)}_{01}-p_i(t)\omega^{(i)}_{10},
\end{equation}
where the rates $\omega^{(i)}_{01}$ and $\omega^{(i)}_{10}$ account to filling up $(0 \rightarrow 1)$ and vice versa $(1 \rightarrow 0)$ respectively. These rates depend on $\epsilon_i$ and $\mu_i$ and are given by
\begin{equation}
\omega^{(i)}_{01}= 
\frac{\Gamma_0}{1+e^{A_i}} \qquad \text{and} \qquad \omega^{(i)}_{10}= 
\frac{\Gamma_0e^{A_i}}{1+e^{A_i}},
\end{equation}
where $\Gamma_0$ quantifies the coupling strength between the system and thermal bath (for simplicity taken equal in all cases) and $A_i=\frac{1}{T}(\epsilon_i-\mu_i)$ for the $i$-th stroke.

Thermodynamic relations are derived starting from the system energy given by $E(t) = \epsilon(t)p(t)$ where we introduce $\epsilon(t)$ / $p(t)$ as the energy level / occupation probability of the QD at time $t$. The time derivative of $E(t)$ gives two contributing terms:
\begin{eqnarray}\label{eqn:edot}
    \dot{E}(t) &=& \;\;\;\;\underbrace{\dot{\epsilon}(t)p(t)}\;\;\;\;+\;\;\;\;\underbrace{\epsilon(t)\dot{p}(t)}\\
    & &\text{"direct drive" + "exchange"} \nonumber
\end{eqnarray}
the former  identified as direct  driving and appearing only during the external driving phases (during which $\dot{\epsilon}(t)\neq 0$), 
and identified as the \emph{direct work} $\dot{W}_{\text{d}}$ given by
\begin{equation}
    \dot{W}_{\text{d}}(t)=\sum_{i=1}^{N} \left(\epsilon_{i+1}-\epsilon_{i}\right)\delta\left(t-\frac{i\tau}{N}\right)p_i(t),
    \label{eqwd}
\end{equation}
where periodic boundary conditions for the index $i$, i.e. $\epsilon_{i+N}=\epsilon_i$ have been  employed. We stress that $\dot{W}_{\text{d}}(t)>0$ according to whether the energy of the QD increases. By averaging it over one full period, one has that
\begin{equation}\label{eqWd1}
\overline{\dot{W}}_{\text{d}}
\equiv\frac{1}{\tau}\int_0^{\tau}\dot{W}_{\text{d}}(t)dt =\frac{1}{\tau}\sum_{i=1}^{N}\left(\epsilon_{i+1}-\epsilon_{i}\right)p_i\left(\frac{i\tau}{N}\right).
\end{equation}
The second term $\epsilon(t)\dot{p}(t)$ appears during the exchange phases and it is different from zero provided an exchange of particles takes place. From 
 the integration of $\dot{p}(t)$ from $\tau_{i-1}=(i-1)\tau/N$ to $\tau_i$, one can re-express it as $\overline{J}_i \equiv dN_i/\tau$ as the average net number of particles exchanged during stage $i$ per period, given by
\begin{equation}\label{J}
    \overline{J}_i = \frac{1}{\tau}\left[p_i\left(\frac{i\tau}{N} \right)-p_i\left(\frac{(i-1)\tau}{N} \right)\right].
\end{equation}
The  \emph{total} energy exchange $dE_i$ during the exchange phase of stage $i$ is rewritten as
\begin{equation}
    dE_i \equiv \int_{\frac{(i-1)\tau}{N}}^{\frac{i\tau}{N}}\epsilon(t)\dot{p_i}(t)dt=\epsilon_i dN_i.
    \label{dei}
\end{equation}
From the reservoir viewpoint, above expression can be split in two parts:
$dE_i=\left(\epsilon_i-\mu_i\right)dN_i+\mu_i dN_i$, the former and latter terms identified as heat and chemical work respectively, counted positive when \emph{delivered to} the QD, i.e. when the energy of the QD increases. 
By averaging Eqs.~(\ref{eqn:edot}) and (\ref{dei})  over one full period and using  Eq.~(\ref{J}), it follows that $\overline{\dot{W}}_{\text{d}}+\overline{\dot{Q}}+\overline{\dot{W}}_{\text{ch}}=0$,
where $\overline{\dot{Q}}$ and $\overline{\dot{W}}_{\text{ch}}$ are given by
\begin{eqnarray}  &&\overline{\dot{Q}}=\sum_{i=1}^{N}\left(\epsilon_i-\mu_i \right)\overline{J}_i\;; \label{eqnQ1}\\   &&\overline{\dot{W}}_{\text{ch}}=\sum_{i=1}^{N}\mu_i \overline{J}_i \;;\label{eqnWchem1}\\
&&\overline{\dot{W}}_{\text{d}}=-\sum_{i=1}^{N}\epsilon_i \overline{J}_i \label{eqnWd1}.
\end{eqnarray}
The entropy production averaged over one full cycle can be determined using Schnakenberg's formula \cite{schnakenberg1976network}:
\begin{equation}
\label{sch}
   \overline{\dot{S}} = \frac {{ k_B}}{\tau} \sum_{i=1}^N \int\limits_{\frac{(i-1)\tau}{N}}^{\frac{i\tau}{N}} [ \omega^{(i)}_{01} - (\omega^{(i)}_{01} + \omega^{(i)}_{10}) { p_i(t)} ]\ln \frac{\omega^{(i)}_{01} {(1- p_i(t))}}{\omega^{(i)}_{10}{ p_i(t)} }\textrm{d} t,
\end{equation}
where from now on we set $k_B=1$. Taking into account the periodicity of $p_i(t)$ and the first law of thermodynamics, this expression can be rewritten in the following form:
\begin{equation}
    \overline{\dot{S}}=-\frac{\overline{\dot{Q}}}{T}=\frac{\overline{\dot{W}}_{\text{d}}+\overline{\dot{W}}_{\text{ch}}}{T},
\end{equation}
consistent to the ratio between the total average heat and the system temperature, as expected.

\subsection{General expressions for the probability distribution and average flux}
Although the QD evolves to the equilibrium regime when it is in contact with only one thermal bath, this is no longer true when it is periodically connected to different reservoirs. From Eq. (\ref{eq:master}) together with the continuity
of $p_i(t)$ at each reservoir switching and taking into account that the system returns to the initial state after a complete period,  one 
obtains the  following (generic) expression
for $p_i(t)$ for the $i$-th stage and $N$ strokes \cite{harunari2020}:
\begin{equation}
	 p_i(t) = p_i^\text{eq} + \left[p \left(\frac{(i-1)\tau}{N}\right) - p_i^\text{eq}\right] e^{-(\omega^{(i)}_{01} + \omega^{(i)}_{10}) (t-\frac{(i-1)\tau}{N})},
\label{pgen}
\end{equation}
where $p_i^{eq}=\omega^{(i)}_{01}/(\omega^{(i)}_{01}+\omega^{(i)}_{10})=(1+e^{\frac{1}{T}(\epsilon_i-\mu_i)})^{-1}$ is obtained
from transition rates $\omega^{(i)}_{01}$ and $\omega^{(i)}_{10}$ at the stage $i$ for $(i-1)\tau/N \leq t < i\tau/N$. By  expressing $p \left((i-1)\tau/N\right)$ in terms
of probabilities from previous strokes, we finally arrive at the generic form for $p_i(t)$:
\begin{multline}
\label{pss}
    p_i(t) = p_i^\text{eq} + e^{-(\omega^{(i)}_{01} + \omega^{(i)}_{10}) (t-\frac{(i-1)\tau}{N})}\left\lbrace\sum_{m=2}^i \xi_{m,i-1} \Delta_{m-1,m}\right.\\ 
    \left.+\frac{\xi_{1,i-1}}{1- \xi_{1,N}} \left[ \Delta_{N,1} + \sum_{n=2}^N \xi_{n,N} \Delta_{n-1,n} \right]  \right\rbrace,
\end{multline}
solely expressed in terms of quantities $\Delta_{i,j} \equiv p_i^\text{eq} - p_j^\text{eq}$ and $\xi_{i,j} \equiv \exp\{ -\frac{\tau}{N} \sum_{n=i}^j (\omega^{(i)}_{01} + \omega^{(i)}_{10}) \}$.
Having $p_i(t)$, average fluxes can be obtained. From Eqs. (\ref{J}) and (\ref{pgen})  the average flux  $\overline{J}_i$ during stage $i$ reads
\begin{multline}
  \overline{J}_i = \frac{1}{\tau} \left\lbrace \frac{\xi_{1,i-1}}{1- \xi_{1,N}} \left[ \Delta_{N,1}+ \sum_{n=2}^N \xi_{n,N} \Delta_{n-1,n} \right] \right. \\
  \left. +\sum_{m=2}^i \xi_{m,i-1} \Delta_{m-1,m} \right\rbrace (\xi_{i,i} - 1).
  \label{genflux}
\end{multline}
From Eq. (\ref{genflux}), the mean flux and thermodynamic quantities can be obtained for a   generic $N$.
\subsection{Two (N=2) and three (N=3) stages collisional engine}
The simplest collisional engine is constituted by $N=2$
strokes, in which at the first stage ($0<t<\tau^*$) the system in contact with the right reservoir (and disconnected from the left reservoir), whereas in the second stage ($\tau^*<t<\tau$) the system is connected to the left reservoir (and disconnected from the right reservoir).
Despite the simplicity, distinct aspects comprising the role of parameters
for  equal \cite{Alexandre17}
and asymmetric switchings and distinct maximization routes \cite{fernando} have been considered. By curbing our analysis to the  the simplest symmetric case $\tau^*=\tau/2$, Eq. (\ref{genflux}) reduces to:
\begin{equation}
    \overline{J_l}=\frac{1}{2\tau}\left[\tanh{\left(\frac{
    A_r}{2}\right)}-\tanh{\left(\frac{
    A_l}{2}\right)}\right]\tanh{\left( \frac{\Gamma_0 \tau}{4}\right)},
    \label{fluxestwo}
\end{equation}
where $\overline{J_r}=-\overline{J_l}$ and indexes $i=r,l$ are
associated with right or left reservoir respectively.
From Eq.~(\ref{eqnQ1}), the total heat exchanged $\overline{\Dot{Q}}=\overline{\Dot{Q}}_r+\overline{\Dot{Q}}_l$ is given by
\begin{equation}    \overline{\Dot{Q}}=(\epsilon_r-\mu_r-\epsilon_l+\mu_l) \overline{J_l},
\end{equation}
whereas chemical work and direct work,   obtained from Eq.~(\ref{eqnWchem1}) and (\ref{eqnWd1}) respectively, read $\overline{\dot{W}}_{\text{chem}}=(\mu_l-\mu_r)\overline{J}_r$ and $\overline{\dot{W}}_{\text{d}}=(\epsilon_r-\epsilon_l)\overline{J}_r$.

Since the main goal of this paper is to tackle the role of  intermediate stages, we present a detailed analysis about the simplest setup with an intermediate stage, namely a cycle composed out of $N=3$ stages. More concretely, the system is placed in contact with the right reservoir during the first third of time, to the middle reservoir in the second stage, and to the left reservoir at the final stage, completing a periodic cycle after $\tau$. From Eq.~(\ref{pgen}), the probability distribution $p_i(t)$ at the $i$-th stage reduces to the following expression
\begin{multline}
    p_i(t)=\frac{1}{e^{A_i}+1}-\frac{e^{\Gamma_0(\frac{i\tau }{3}-t)}}{\phi} \left(\frac{1+e^{\frac{\Gamma_0\tau}{3}}}{e^{A_i}+1}\right.\\\left.-\frac{1}{e^{A_{i+1}}+1}-\frac{e^{\frac{\Gamma_0\tau}{3}}}{e^{A_{i+2}}+1}\right)
\end{multline}
respectively, where $\phi=1+e^{\frac{\Gamma_0\tau}{3}}+e^{\frac{2\Gamma_0\tau}{3}}\geq3$. From Eq. (\ref{genflux}) for $N=3$, each mean flux $\overline{J_i}$ reduces to the following expression: 
\begin{multline}
    \overline{J_i}=\Omega \bigg[\left(1+e^{A_{i+1}}\right) \left(e^{A_i}-e^{A_{i-1}}\right) e^{\frac{\Gamma_0\tau}{3}}\\ 
    +\left(e^{A_{i-1}}+1\right) \left(e^{A_i}-e^{A_{i+1}}\right)\bigg],
    \label{fluxeth}
\end{multline}
where 
\begin{equation}
    \Omega = \frac{e^{\frac{\Gamma_0\tau}{3}}-1}{(e^{A_r}+1)(e^{A_m}+1)(e^{A_l}+1)\phi \tau}>0 .
\end{equation}
 Quantities $\overline{\dot{W}}_{\text{chem}}$  and $\overline{\dot{W}}_{\text{d}}$ are straightforwardly obtained from Eqs. (\ref{eqnWchem1}), (\ref{eqnWd1}) and (\ref{fluxeth}), given by 
$\overline{\dot{W}}_{\text{chem}}=(\mu_l-\mu_m){\overline J}_l+(\mu_r-\mu_m){\overline J}_r$ and 
$\overline{\dot{W}}_{\text{d}}=(\epsilon_m-\epsilon_l){\overline J}_l+(\epsilon_m-\epsilon_r){\overline J}_r$,
respectively.

\begin{figure}[t]
	\centering
    \includegraphics[width=0.85\linewidth]{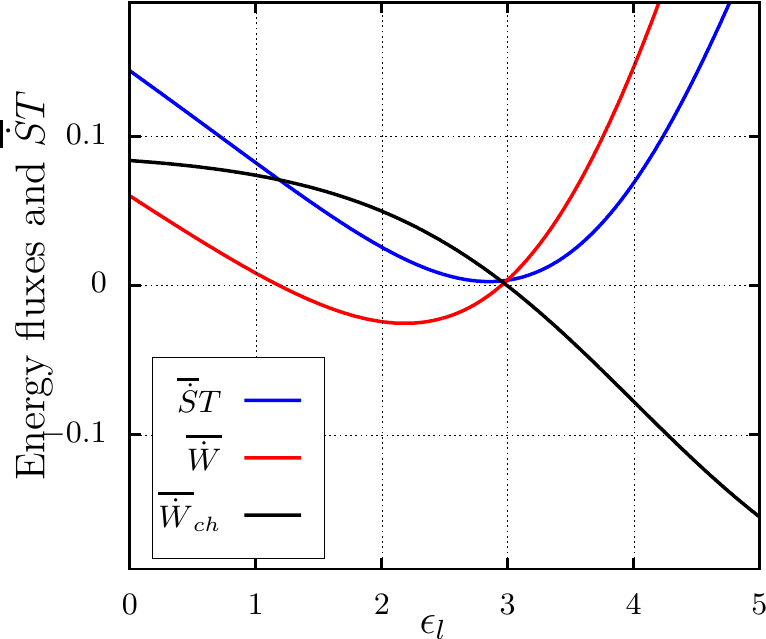}
	 \caption{For three stages case, the entropy production times temperature $\overline{\dot{S}}T$ (blue), direct work $\overline{\dot{W}}_{\text{d}}$ (red) and chemical work  $\overline{\dot{W}}_{\text{ch}}$ (black) vs $\epsilon_l$. Parameters:  $T=1$, $\tau=0.1$, $\Gamma_0=1$, $\epsilon_r=1$, $\mu_r=2$, $\epsilon_m=1.7$, $\mu_m=3$ and $\mu_l=4$. }
	\label{Q_W3}
\end{figure}

\begin{figure}[t]
	\centering
        \includegraphics[width=0.85\linewidth]{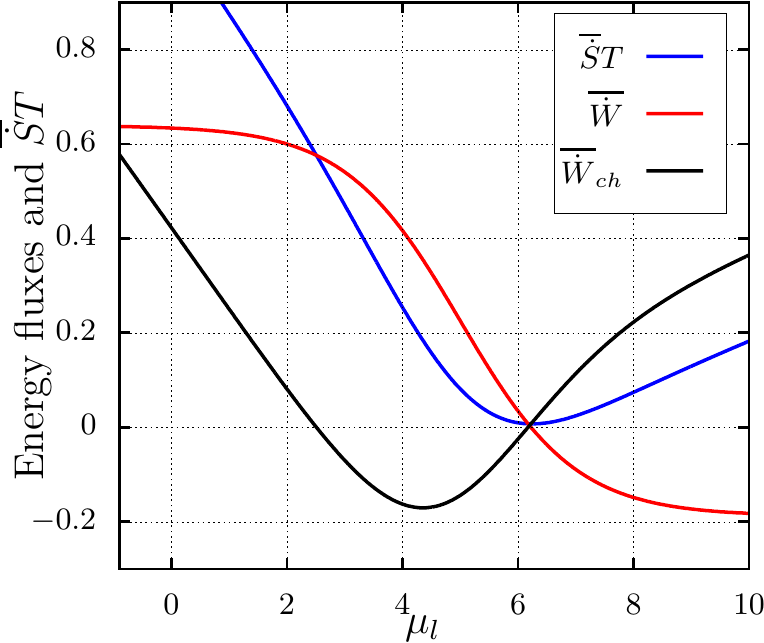}
	 \caption{For three stages case, the entropy production times temperature $\overline{\dot{S}}T$ (blue), direct work $\overline{\dot{W}}_{\text{d}}$ (red) and chemical work  $\overline{\dot{W}}_{\text{ch}}$ (black) vs $\mu_l$. Parameters: $T=1$, $\tau=0.1$, $\Gamma_0=1$, $\epsilon_r=1$, $\mu_r=2$, $\epsilon_m=1.7$, $\mu_m=3$ and $\epsilon_l=5$. }
	\label{Q_W3_vs_mu}
\end{figure}

A first insight into the behavior of thermodynamic quantities ($T\overline{\dot{S}},\ \overline{\dot{W}}_{\text{d}}$ and $\ \overline{\dot{W}}_{\text{ch}}$) is depicted in
Figs. \ref{Q_W3} and  \ref{Q_W3_vs_mu} upon varying the energy $\epsilon_l$
and chemical potential $\mu_l$ respectively. In the former case, there is a closed region  delimited by approximately  $1<\epsilon_l<3$ in which $\ \overline{\dot{W}}_{\text{d}}<0$ and $\ \overline{\dot{W}}_{\text{ch}}>0$, consistent to an engine operation (as shall be explained further). Conversely,
for $\epsilon_l>3$, one has that   $\ \overline{\dot{W}}_{\text{d}}>0$ and $\ \overline{\dot{W}}_{\text{ch}}<0$, consistent to a pump regime operation.
Similar findings are depicted in Fig. \ref{Q_W3_vs_mu}, but as $\mu_l$ is varied the pump regime is delimited by a closed region $2.5<\mu_l<6$, whereas
the engine mode operation extends to $\mu_l>6.2$.

\section{Efficiency}
\begin{figure}[t]
	\centering
    \includegraphics[width=0.85\linewidth]{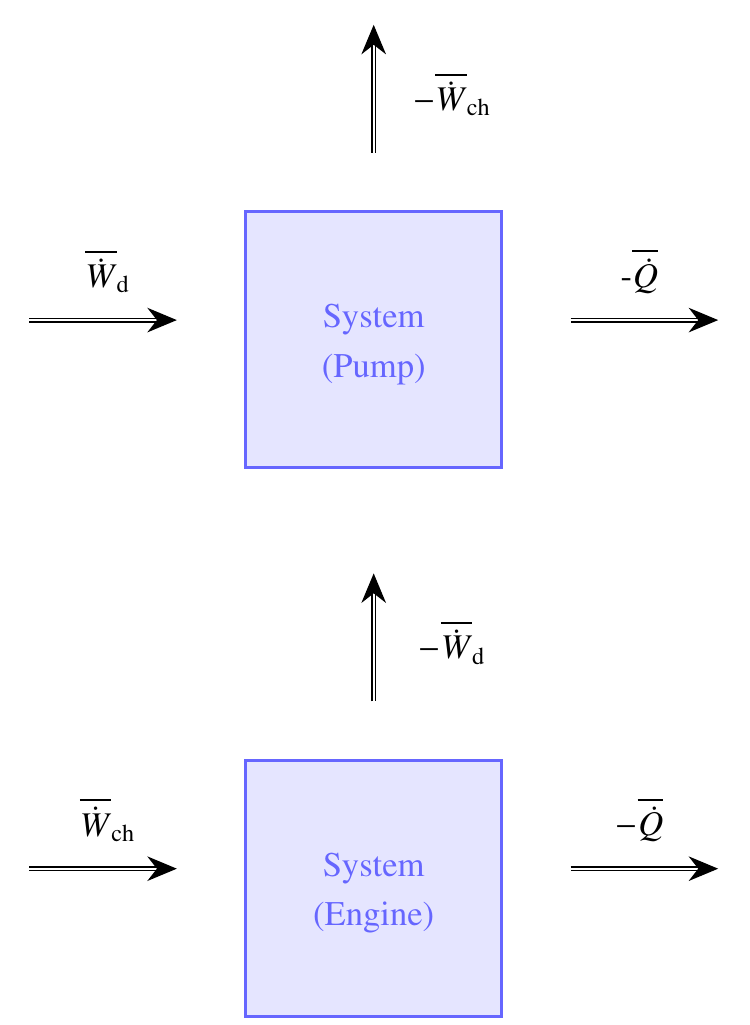}
	 \caption{Sketch of a pump (top) and engine (bottom) working operation}. 
	\label{sketch}
\end{figure}

Having introduced the main features of the model and obtained expressions for the relevant thermodynamical quantities, we are now in position to describe  the operation of our collisional
system as engine or as pump, together with the existence of distinct optimization routes. To start, a short comment about the sign of thermodynamic quantities in each regime is useful in order to establish a reliable definition of the efficiency. A particle pump typically consumes direct work $\overline{\dot{W}}_{\text{d}}$ in order to move a particle from a lower to a higher chemical potential, which is consistent with $\ \overline{\dot{W}}_{\text{ch}}<0$ and $\overline{\dot{W}}_{\text{d}}>0$ (see e.g. top panel of Fig. \ref{sketch}). Since there is no particle accumulation in the QD, such delivered chemical work can only be the result of transferring particles from reservoirs at lower to those to higher chemical potentials.  A reliable definition of power is given by ${\cal P}_{\text{\tiny pump}}=-\overline{\dot{W}}_{\text{ch}}$ with associate efficiency $\eta_{\text{pump}}$ given by
\begin{equation}\label{eta}
        \eta_{\text{pump}}=\frac{{\cal P}_{\text{\tiny pump}}}{\overline{\dot{W}}_{\text{d}} },
\end{equation}
respectively.  By construction, above efficiency definition implies that  $0\le \eta_{\text{pump}}\le 1$.

With an appropriate choice of parameters the  system can also operate as an engine ($\overline{\dot{W}}_{\text{d}}<0$ and $\ \overline{\dot{W}}_{\text{ch}}>0$), as sketched in bottom panel of Fig. \ref{sketch}. The dynamics are similar, but in this case the power output
is given by ${\cal P}_{\text{\tiny engine}}=-\overline{\dot{W}}_{\text{d}}\ge 0$ according to the efficiency definition:
  \begin{equation}\label{etaww}
    \eta_{\text{engine}}=\frac{{\cal P}_{\text{\tiny engine}}}{\overline{\dot{W}}_{\text{ch}} } ,
\end{equation}
where $0\le \eta_{\text{engine}}\le 1$. It is worth pointing out that such efficiency definitions state that $\eta_{\text{pump}}=1/\eta_{\text{engine}}$. For the two simplest
$N=2$ and $N=3$ cases,
efficiencies are given by 
\begin{equation}
\eta_{\text{engine},N=2}=(\epsilon_l-\epsilon_r)/(\mu_l-\mu_r),
\end{equation}
and 
\begin{equation}
\eta_{\text{engine},N=3}=\frac{(\epsilon_l-\epsilon_m){\overline J}_l+(\epsilon_r-\epsilon_m){\overline J}_r}{(\mu_m-\mu_l){\overline J}_l+(\mu_m-\mu_r){\overline J}_r},
\end{equation}
respectively. While the former solely depends on $\epsilon_l,\epsilon_r,\mu_l$ and $\mu_r$, due to the fact that ${\overline J}_l=-{\overline J}_r$, the latter is more
revealing and efficiencies  depend on
the interplay with intermediate parameters ($\mu_m,\epsilon_m$) and the period $\tau$.
\begin{figure}[t]
	\centering
        \includegraphics[width=0.85\linewidth]{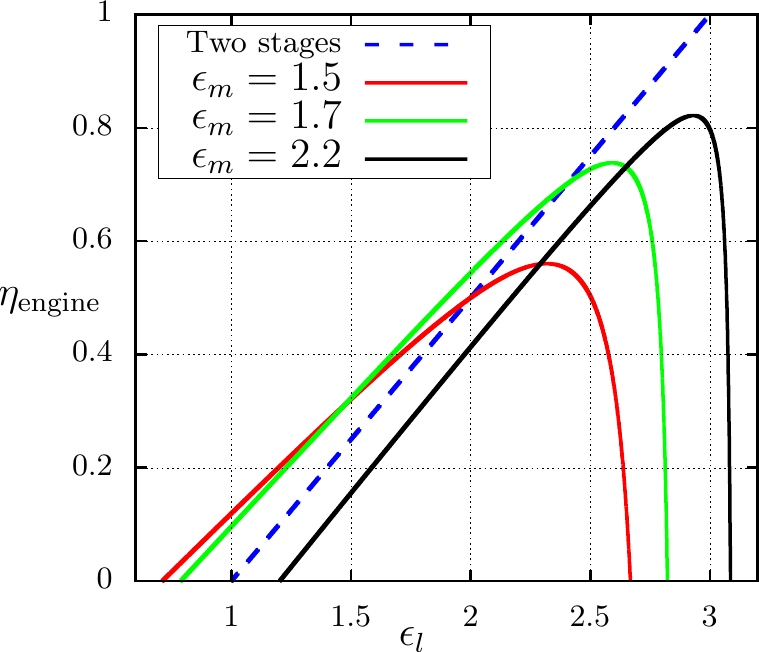}
        \includegraphics[width=0.85\linewidth]{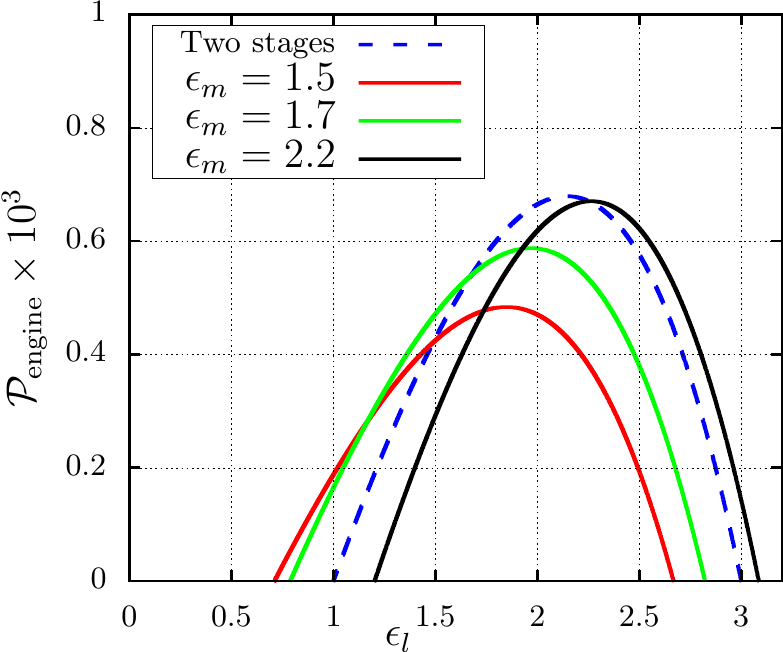}
	\caption{ For the engine operation mode and distinct $\epsilon_m$'s, the depiction of efficiency $\eta$ (top) and power ${\cal P}_{engine}$ (bottom) versus  
	$\epsilon_l$ for $N=2$ and $N=3$ stages. Parameters: $\tau=15$, $\Gamma_0=1$, $\epsilon_r=1$, $\mu_r=2$, $\mu_m=3$ and $\mu_l=4$. }
	\label{com}
\end{figure}
Analogous expressions are  straightforwardly obtained for $N>3$. Giving that
they are longer (and less instructive), they will be omitted here. 
\begin{figure}[t]
	\centering
        \includegraphics[width=0.85\linewidth]{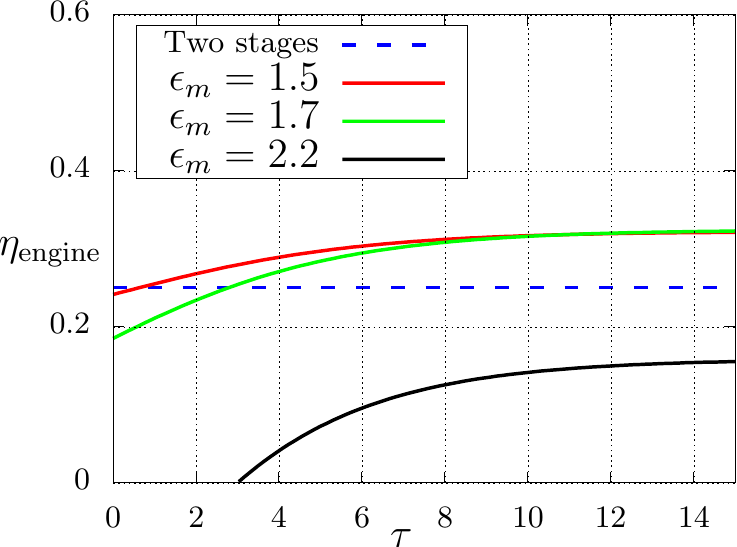}
	\caption{For  $\epsilon_l$ fixed and distinct $\epsilon_m$'s, the depiction of efficiency $\eta$ versus period $\tau$ for $N=2$ and $N=3$. Parameters:  $\Gamma_0=1$, $\epsilon_r=1$, $\epsilon_l=1.5$, $\mu_r=2$, $\mu_m=3$ and $\mu_l=4$.  }
	\label{etat}
\end{figure}
\subsection{Influence of intermediate reservoirs}
A first insight about the influence of model parameters ($N, \epsilon_i$'s, $\mu_i$'s and $\tau$) is  depicted in  Figs. \ref{com} and \ref{etat} for three different values of $\epsilon_m$ by varying $\epsilon_l$ and $\tau$ respectively, and  the parameters  $\epsilon_r$, $\mu_r$, $\mu_m$ and $\mu_l$ held fixed (for simplicity). In the former case (fixed $\tau$), the system efficiency is solely improved   via adjustment of $\epsilon_l$ (or $\epsilon_r$) and hence $\eta$
exhibits a linear dependence on $\epsilon_l$ for $N=2$, 
reaching the maximum (ideal) limit  $\eta_{ME}=1$ with ${\cal P}_{ME}=0$, consistent with the reversible operation mode. Conversely, 
distinct routes for optimization become feasible for $N=3$, such as by varying $\epsilon_m$ and $\epsilon_l$ and as a consequence, there are some regions in
the set of parameters in which the inclusion of stages can confer a larger
performance.  However, maximum efficiency $\eta_{ME}$ and  power ${\cal P}_{M{\cal P}}$ for $N=2$ is somewhat superior than for $N=3$. Likewise, Fig. \ref{etat} reinforces such advantages for $N=3$ by illustrating that efficiencies $\eta$'s  become superior as the duration $\tau$ of cycle increases.

Fig. \ref{com_mult_eff_pow} extends the above analysis by tackling number of strokes for $N$ ranging from $2$ to $5$ by depicting $\eta$ and ${\cal P}$ for distinct values of $\epsilon_l$ and $\epsilon_r,\mu_r,\mu_l$ held fixed. Due to the existence of several parameters, we adopt a simple criterion for choosing intermediate $\mu_m$'s,  by changing them by a fixed amount $\mu_m=\mu_{m-1}+\Delta\mu$, where $\Delta\mu=(\mu_l - \mu_r )/N$
and corresponding $\epsilon_m$'s as proportional to them $\epsilon_m=\alpha \mu_m$  for all (intermediate) $m$. As for $N=3$, the system performance can be improved by suitable choice of parameters and properly increasing the  number of stages. Also like $N=3$, associate maximum power and efficiencies are somewhat lower than for $N=2$. 

\subsection{Optimal parameter choices and maximizations for $N=3$ stages }
\begin{figure}[t!]
	\centering
        \includegraphics[width=0.85\linewidth]{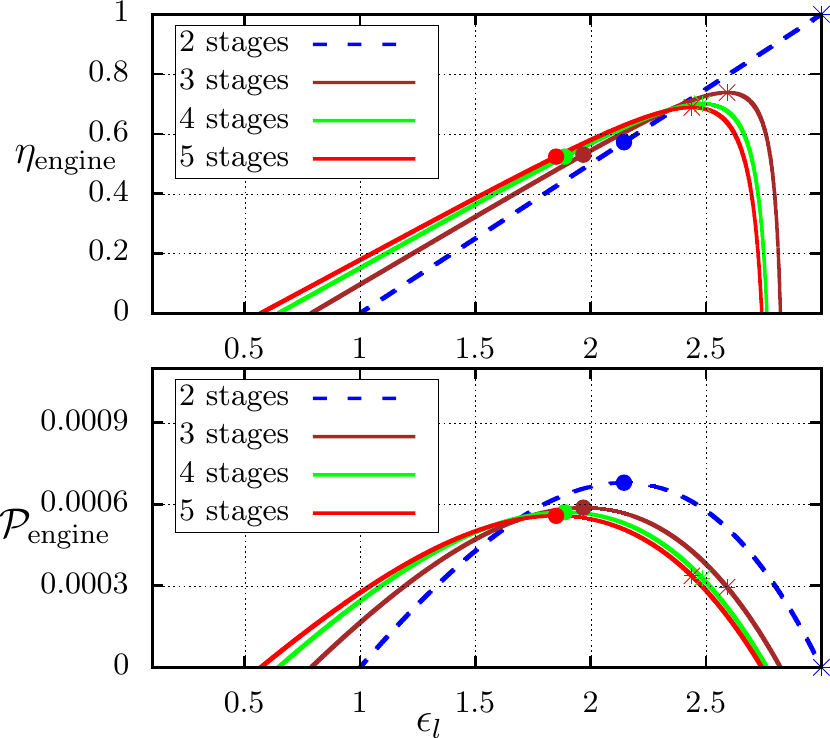}
	\caption{ For the engine operation mode, the comparison between the efficiencies $\eta$'s (top) and power outputs ${\cal P}$'s (bottom) for the two stages system (blue) and the three stages system (brown), four stages (green) and five stages (red) at engine regime vs. $\epsilon_l$. Asterisks and circles show the points where efficiency and power have a maximum values respectively. Parameters: $\alpha=0.56$, $\tau=15$, $\Gamma_0=1$, $\epsilon_r=1$, $\mu_r=2$ and $\mu_l=4$. }
	\label{com_mult_eff_pow}
\end{figure}

Aformentioned results indicate that the increase of stages may improve the
system performance. Here we exploit the engine performance for distinct sets of intermediate parameters and operation modes. Since Fig. \ref{com_mult_eff_pow} indicate a common set of similar findings for $N>2$, our analysis will be carried
out for $N=3$ for the following choice of parameters:  $(\epsilon_r,\mu_r)$ always held fixed and by varying $(\epsilon_l,\epsilon_m)$  [for $(\mu_l,\mu_m)$  fixed] and the other way around. Despite the exactness of all results, expressions for maximized quantities  are quite cumbersome and have to be found numerically by solving  transcendental equations. In particular, maximal powers ${\cal P}_{M{\cal P}}$'s yield at $\epsilon^*_m$ (fixed $\epsilon_l$) and $\epsilon^*_l$ (for fixed $\epsilon_m$) and obey the following expressions:
\begin{equation}
{\overline J}_r(\epsilon^*_m)+{\overline J}_l(\epsilon^*_m)=(\epsilon_l-\epsilon^*_m){\overline J}'_l(\epsilon^*_m)+(\epsilon_r-\epsilon^*_m){\overline J}'_r(\epsilon^*_m),
\label{m1}
\end{equation}
in the former case ($\epsilon_l$ fixed), with maximum power 
\begin{equation}
    {\cal P}_{M{\cal P}}=(\epsilon_l-\epsilon^*_m){\overline J}_l(\epsilon^*_m)+(\epsilon_r-\epsilon_m){\overline J}_r(\epsilon^*_m)
\end{equation}
and 
\begin{equation}
{\overline J}_l(\epsilon^*_l)=\frac{1}{\epsilon_m}[\epsilon^*_l{\overline J}'_l(\epsilon^*_l)+(\epsilon_r-\epsilon_m){\overline J}'_r(\epsilon^*_l)],
\label{m2}
\end{equation}
in the latter case (for fixed $\epsilon_m$) with maximum power ${\cal P}_{M{\cal P}}=(\epsilon^*_l-\epsilon_m){\overline J}_l(\epsilon^*_l)+(\epsilon_r-\epsilon_m){\overline J}_r(\epsilon^*_l)$.  Here  ${\overline J}'_X(\epsilon^*_Y)\equiv\partial {\overline J}_X/\partial \epsilon_X$ evaluated at $\epsilon^*_Y$ and the notation ${\overline J}_Y(\epsilon_X)$ has been adopted here in order to state which energy was taken into account in the calculation of derivative.
 Remarkably, the locus of maximum $\epsilon^*_m$ (fixed $\epsilon_l$) and  $\epsilon^*_l$ (fixed $\epsilon_m$)
intersect at a single point (${\bar \epsilon}_m,{\bar \epsilon}_l$),  marking the  existence of a global maximization of power, given by ${\cal P}^*_{M{\cal P}}=({\bar \epsilon}_l-{\bar \epsilon}_m){\overline J}_l({\bar \epsilon_l})+({\bar \epsilon}_r-{\bar \epsilon}_m){\overline J}_r({\bar \epsilon}_l)$. Analogous 
expressions are obtained for the pump regime operation just by replacing $\epsilon_X\rightarrow \mu_X$.

\begin{figure*}[t]
	\centering
        \includegraphics[width=0.4\linewidth]{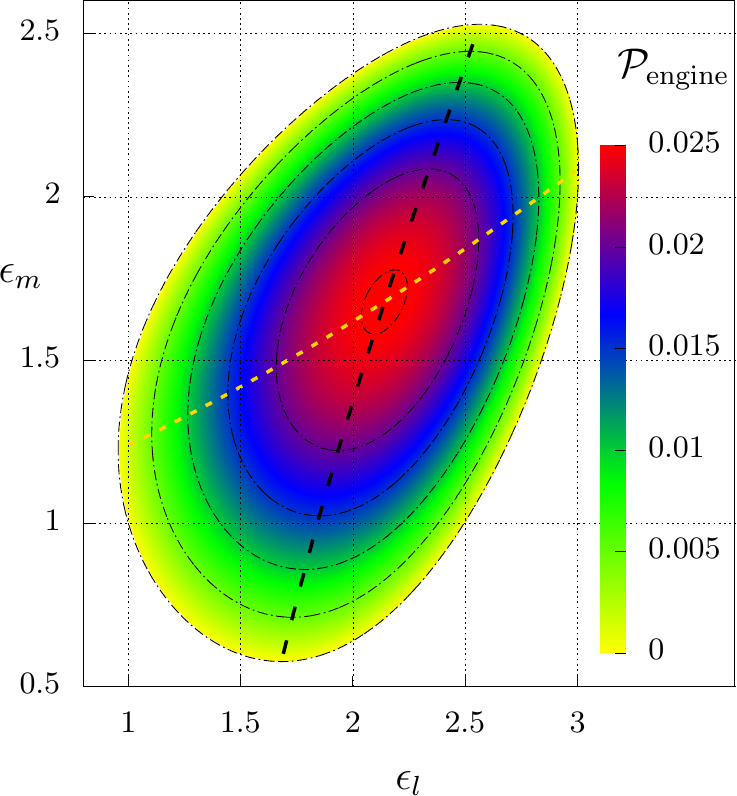}
        \hspace{5mm}
        \includegraphics[width=0.4\linewidth]{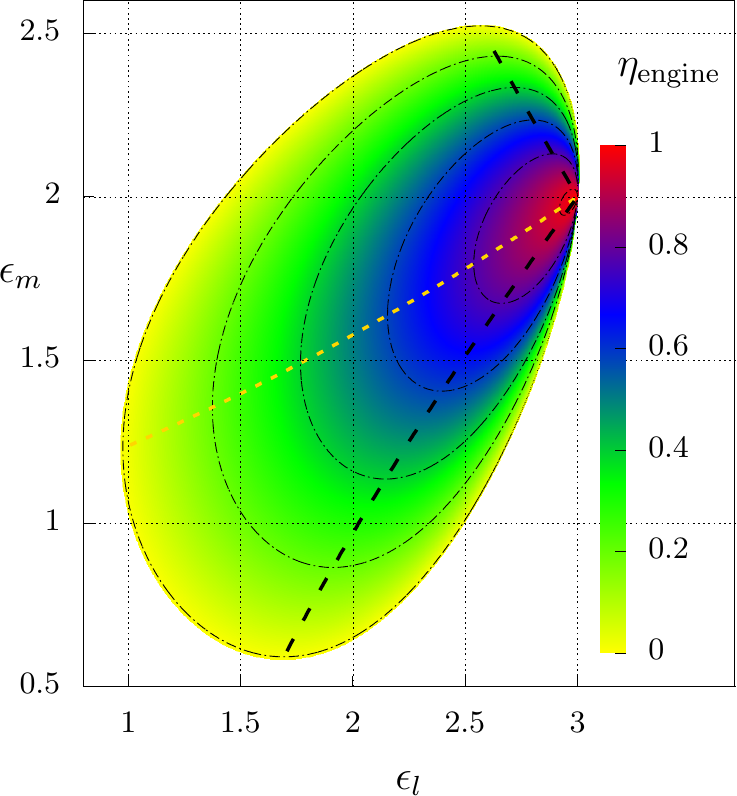}
	\caption{ For the engine operation mode and $N=3$, the power and efficiency heat maps  for several values of $\epsilon_l$ and $\epsilon_m$. White zones correspond to the dud  regime. Parameters: $\tau=1$, $\Gamma_0=1$, $T=1$, $\epsilon_r=1$, $\mu_r=2$, $\mu_m=3$ and $\mu_l=4$. }
	\label{3Dpo}
\end{figure*}

\begin{figure*}[t]
	\centering
        \includegraphics[width=0.4\linewidth]{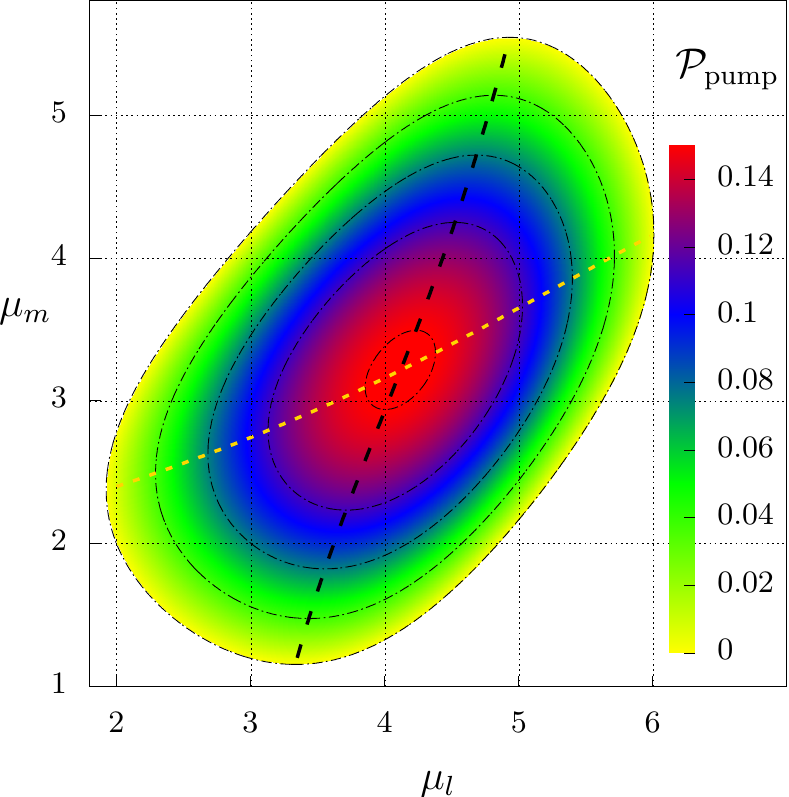}
        \hspace{5mm}
        \includegraphics[width=0.4\linewidth]{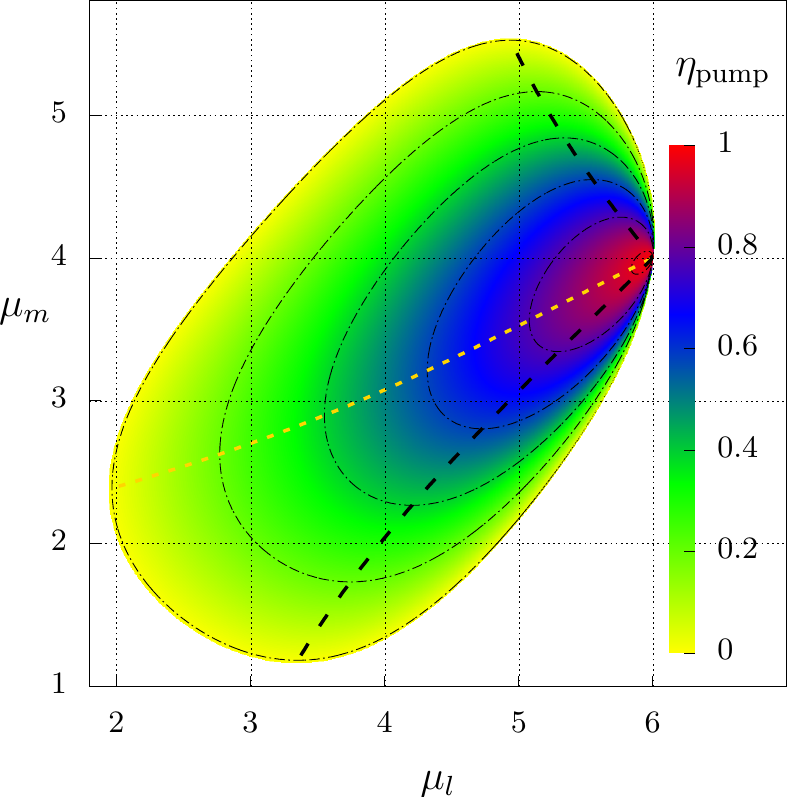}
	\caption{  For the pump operation mode and $N=3$, the power and efficiency heat maps for several values of $\mu_l$ and $\mu_m$. White zones corresponds to the dud  regime. Parameters: $\tau=1$, $\Gamma_0=1$, $T=1$, $\epsilon_r=1$, $\mu_r=2$, $\epsilon_m=3$ and $\epsilon_l=5$. }
	\label{3Dpopump}
\end{figure*}

Figs. \ref{3Dpo} and \ref{3Dpopump}  summarize our main findings for engine and
pump regimes, respectively. Firstly, both  ${\cal P}$ and $\eta$ and be optimized under  suitable choice of intermediate parameters $\epsilon_m^*$ and $\mu_m^*$  (dashed lines fulfill Eq.~(\ref{m1})), the former  also
being simultaneously maximized with respect to $\epsilon_l$ (or $\mu_l$), characterized by a central region 
and given by the intersection between above  lines.
Efficiency heat maps are similar to power-output ones, but regions of larger efficiencies are shifted to larger values of $\epsilon_l/\mu_l$. Unlike ${\cal P}$,  maximum efficiency lines do not
intersect but they merge  at the ideal limit operation $\eta\rightarrow 1$, consistent
with zero dissipation $\overline{\dot{S}}=0$. 
Once again, it is instructive to compare  $N=2$ and $N=3$ efficiencies for  the same sets of  parameters $({\epsilon_l,\epsilon_r,\mu_l,\mu_r})$.  While   optimal $\epsilon^*_m$'s ($\epsilon_l$ held fixed)  provide  larger efficiencies (dotted lines)  than for $N=2$ when $\epsilon_l$ is small, $\eta_{MP}$'s approach to each other as $\epsilon_l$'s increases. Similar findings has been verified for the pump regime in Fig.~\ref{3Dpopump} when $\mu_l$ is small and large.
We close this section by drawing a comparison between  heat maps from Fig.~\ref{3Dpo} for the opposite case, $\mu_r=4>\mu_l=2$ (not shown).
Although heat maps are akin,  optimal power and efficiency regions move for lower sets of $(\epsilon_l,\epsilon_m$), including negative values. The ideal operation regime yields at $\epsilon_l=-1$ and 
$\epsilon_m\approx 0$. Conversely, the global maximum power  ${\cal P}^*_{M{\cal P}}$ in such case is approximately half from that in Fig. \ref{3Dpo}.

\section{Conclusions}
 
We addressed an alternative route for optimizing the  performance of a collisional nonequilibrium setup composed of a quantum-dot,  exposed
sequentially with distinct reservoirs. 
Quantum-dot engines have been broadly investigated in the realm of nonequilibrium and quantum thermodynamics and the present system
yields an exact solution, irrespective of its projection 
(energy, chemical potential, period and number of strokes). Despite this, its behavior is rich enough for exhibiting distinct operations, such as a work-to-work transducer, engine, heat engine and pump. The main point concerning our findings is that a suitable choice of number of stages can provide a better system performance than fewer strokes, above all when the period $\tau$ is longer (see e.g. Fig.~\ref{com}). Thereby, a better performance of QD setups for large $N$ and $\tau$ might be not only more feasible but also more advantageous from the experimental point of view.  It is instructive to compare  with collisional Brownian engines \cite{stable2020thermodynamics,noa2021efficient}, which operate less efficiently for large $\tau$. As potential perspectives, we highlight to extend the idea of splitting the dynamics into distinct strokes for other engine setups, such as those presenting collective effects \cite{gatien,forao2023powerful,bettmann2023thermodynamics}.

\begin{acknowledgements}
The financial support from Brazilian agencies CNPq, CAPES and FAPESP (under grant 2021/03372-2) is
acknowledged.  This project was supported by the Special Research Fund (BOF) of Hasselt University under Grant No. BOF20BL17. 
\end{acknowledgements}

\bibliography{refs}

\end{document}